\documentclass[sigconf]{acmart}
 \usepackage{amsmath,amssymb,amsfonts,bm}
\usepackage{graphicx}
\usepackage{textcomp}
\usepackage{xcolor}
\graphicspath{ {pic/} }
\usepackage{float}
\usepackage{array}
\usepackage{tabularx}
\usepackage{array}
\usepackage{balance}
\usepackage{marvosym}
\usepackage{multirow}
\usepackage{array}
\usepackage{enumitem}
\usepackage{booktabs}
\usepackage{algpseudocode}
\usepackage[ruled]{algorithm2e} 
\usepackage{xcolor}
\usepackage{pifont}
\usepackage{url}

\setcopyright{iw3c2w3} 
\AtBeginDocument{%
	\providecommand\BibTeX{{%
			\normalfont B\kern-0.5em{\scshape i\kern-0.25em b}\kern-0.8em\TeX}}}

\copyrightyear{2021}
\acmYear{2021} 
\acmConference[WWW '21]{Proceedings of the Web Conference 2021}{April 19--23,
	2021}{Ljubljana, Slovenia}
\acmBooktitle{Proceedings of the Web Conference 2021 (WWW '21), April 19--23, 2021,
	Ljubljana, Slovenia}
\acmPrice{}
\acmDOI{10.1145/3442381.3449844}
\acmISBN{978-1-4503-8312-7/21/04}

\makeatletter
\def\@copyrightspace{\relax}
\makeatother

\begin{document}
	
	\title{Self-Supervised Multi-Channel Hypergraph Convolutional Network for Social Recommendation}

	\author{Junliang Yu}
	\affiliation{%
	  \institution{The University of Queensland}}
	\email{jl.yu@uq.edu.au}
	
	\author{Hongzhi Yin}
	\authornote{Corresponding author and having equal contribution with the first author.}
	\affiliation{%
	\institution{The University of Queensland}}
	\email{h.yin1@uq.edu.au}
	
	\author{Jundong Li}
	\affiliation{%
	 \institution{University of Virginia}}
	\email{jundong@virginia.edu}
	
	\author{Qinyong Wang}
	\affiliation{%
	\institution{The University of Queensland}}
	\email{qinyong.wang@uq.edu.au}
	
	\author{Nguyen Quoc Viet Hung}
	\affiliation{%
	  \institution{Griffith University}}
	\email{quocviethung1@gmail.com}
	
	\author{Xiangliang Zhang}
	\affiliation{%
	  \institution{King Abdullah University of Science and Technology}}
	\email{xiangliang.zhang@kaust.edu.sa}

	\renewcommand{\shortauthors}{Yu and Yin, et al. }
	
	\begin{abstract}
		Social relations are often used to improve recommendation quality when user-item interaction data is sparse in recommender systems. Most existing social recommendation models exploit pairwise relations to mine potential user preferences. However, real-life interactions among users are very complex and user relations can be high-order. Hypergraph provides a natural way to model high-order relations, while its potentials for improving social recommendation are under-explored. In this paper, we fill this gap and propose a multi-channel hypergraph convolutional network to enhance social recommendation by leveraging high-order user relations. Technically, each channel in the network encodes a hypergraph that depicts a common high-order user relation pattern via hypergraph convolution. By aggregating the embeddings learned through multiple channels, we obtain comprehensive user representations to generate recommendation results. However, the aggregation operation might also obscure the inherent characteristics of different types of high-order connectivity information. To compensate for the aggregating loss, we innovatively integrate self-supervised learning into the training of the hypergraph convolutional network to regain the connectivity information with hierarchical mutual information maximization. Extensive experiments on multiple real-world datasets demonstrate the superiority of the proposed model over the current SOTA methods, and the ablation study verifies the effectiveness and rationale of the multi-channel setting and the self-supervised task. The implementation of our model is available at \url{https://github.com/Coder-Yu/QRec}.
	\end{abstract}


	
	\maketitle

	\begin{figure}[th]
	\centering
	\includegraphics[width=0.45\textwidth]{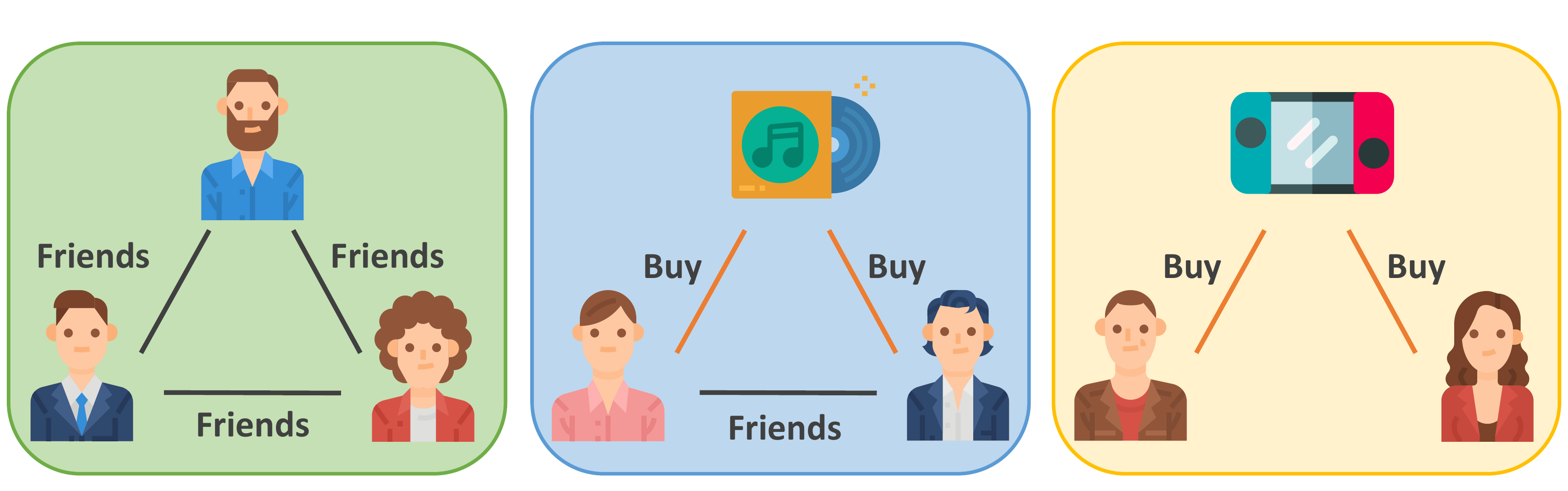}
	\caption{The common types of high-order user relations in social recommender systems. }
	\label{figure.1}
	\vspace{-10px}		
\end{figure}

	\section{Introduction}
	Over the past decade, the social media boom has dramatically changed people's ways of thinking and behaving. It has been revealed that people may alter their attitudes and behaviors in response to what they perceive their friends might do or think, which is known as the \emph{social influence} \cite{cialdini2004social}. Meanwhile, there are also studies \cite{Mcpherson2001Birds} showing that people tend to build connections with others who have similar preferences with them, which is called the \emph{homophily}. Based on these findings, social relations are often integrated into recommender systems to mitigate the data sparsity issue \cite{sarwar2001sparsity,guo2019streaming}. Generally, in a social recommender system, if a user has few interactions with items, the system would rely on her friends' interactions to infer her preference and generate better recommendations. Upon this paradigm, a large number of social recommendation models have been developed \cite{Ma2009Learning2,ma2011recommender,zhao2014leveraging,Guo2015TrustSVD,yu2018adaptive,yu2019generating} and have shown stronger performance compared with general recommendation models. \par
	Recently, graph neural networks (GNNs) \cite{wu2020comprehensive} have achieved great success in a wide range of areas. Owing to their powerful capability in modeling relational data, GNNs-based models also have shown prominent performance in social recommendation \cite{wu2019neural,yu2020enhance,fan2019graph,wu2020diffnet++,wu2019dual,liu2020heterogeneous}. However, a key limitation of these GNNs-based social recommendation models is that they only exploit the simple pairwise user relations and ignore the ubiquitous high-order relations among users. Although the long-range dependencies of relations (i.e. transitivity of friendship), which are also considered high-order, can be captured by using \emph{k} graph neural layers to incorporate features from \emph{k}-hop social neighbors, these GNNs-based models are unable to formulate and capture the complex high-order user relation patterns (as shown in Fig. 1) beyond pairwise relations. For example, it is natural to think that two users who are socially connected and also purchased the same item have a stronger relationship than those who are only socially connected, whereas the common purchase information in the former is often neglected in previous social recommendation models. \par
	
	Hypergraph \cite{bretto2013hypergraph}, which generalizes the concept of edge to make it connect more than two nodes, provides a natural way to model complex high-order relations among users. Despite the great advantages over the simple graph in user modeling, the strengths of hypergraph are under-explored in social recommendation. In this paper, we fill this gap by investigating the potentials of fusing hypergraph modeling and graph convolutional networks, and propose a \textbf{M}ulti-channel \textbf{H}ypergraph \textbf{C}onvolutional \textbf{N}etwork (\textbf{MHCN}) to enhance social recommendation by exploiting high-order user relations. Technically, we construct hypergraphs by unifying nodes that form specific triangular relations, which are instances of a set of carefully designed triangular motifs with underlying semantics (shown in Fig. 2). As we define multiple categories of motifs which concretize different types of high-order relations such as `having a mutual friend', `friends purchasing the same item', and `strangers but purchasing the same item' in social recommender systems, each channel of the proposed hypergraph convolutional network undertakes the task of encoding a different motif-induced hypergraph. By aggregating multiple user embeddings learned through multiple channels, we can obtain the comprehensive user representations which are considered to contain multiple types of high-order relation information and have the great potentials to generate better recommendation results with the item embeddings. \par
	However, despite the benefits of the multi-channel setting, the aggregation operation might also obscure the inherent characteristics of different types of high-order connectivity information \cite{yu2014mixed}, as different channels would learn embeddings with varying distributions on different hypergraphs. To address this issue and fully inherit the rich information in the hypergraphs, we innovatively integrate a self-supervised task \cite{hjelm2018learning,velickovic2019deep} into the training of the multi-channel hypergraph convolutional network. Unlike existing studies which enforce perturbations on graphs to augment the ground-truth \cite{you2020graph}, we propose to construct self-supervision signals by exploiting the hypergraph structures, with the intuition that the comprehensive user representation should reflect the user node's local and global high-order connectivity patterns in different hypergraphs. Concretely, we leverage the hierarchy in the hypergraph structures and hierarchically maximizes the mutual information between representations of the user, the user-centered sub-hypergraph, and the global hypergraph. The mutual information here measures the structural informativeness of the sub- and the whole hypergraph towards inferring the user features through the reduction in local and global structure uncertainty. Finally, we unify the recommendation task and the self-supervised task under a \textit{primary \& auxiliary} learning framework. By jointly optimizing the two tasks and leveraging the interplay of all the components, the performance of the recommendation task achieves significant gains.\par
    The major contributions of this paper are summarized as follows:
	\begin{itemize}[leftmargin=*]
		\item We investigate the potentials of fusing hypergraph modeling and graph neural networks in social recommendation by exploiting multiple types of high-order user relations under a multi-channel setting. 
		\item We innovatively integrate self-supervised learning into the training of the hypergraph convolutional network and show that a self-supervised auxiliary task can significantly improve the social recommendation task.
		\item We conduct extensive experiments on multiple real-world datasets to demonstrate the superiority of the proposed model and thoroughly ablate the model to investigate the effectiveness of each component with an ablation study.
	\end{itemize} 
	
	\section{Related Work}
	\subsection{Social Recommendation}
	As suggested by the social science theories \cite{cialdini2004social,Mcpherson2001Birds}, users' preferences and decisions are often influenced by their friends. Based on this fact, social relations are integrated into recommender systems to alleviate the issue of data sparsity. Early exploration of social recommender systems mostly focuses on matrix factorization (MF), which has a nice probabilistic interpretation with Gaussian prior and is the most used technique in social recommendation regime. The extensive use of MF marks a new phase in the research of recommender systems. A multitude of studies employ MF as their basic model to exploit social relations since it is very flexible for MF to incorporate prior knowledge. The common ideas of MF-based social recommendation algorithms can be categorized into three groups: co-factorization methods \cite{ma2008sorec,yang2017social}, ensemble methods \cite{Ma2009Learning}, and regularization methods \cite{ma2011recommender}. Besides, there are also studies using socially-aware MF to model point-of-interest \cite{yin2016discovering,Yin2016Adapting,yin2018joint}, preference evolution \cite{wu2017modeling}, item ranking \cite{zhao2014leveraging,yu2018adaptive}, and relation generation \cite{yu2019generating,gao2020recommender}. 
	\par
	Over the recent years, the boom of deep learning has broadened the ways to explore social recommendation. Many research efforts demonstrate that deep neural models are more capable of capturing high-level latent preferences \cite{yin2019social,yin2020overcoming}. Specifically, graph neural networks (GNNs) \cite{zhou2018graph} have achieved great success in this area, owing to their strong capability to model graph data. GraphRec \cite{fan2019graph} is the first to introduce GNNs to social recommendation by modeling the user-item and user-user interactions as graph data. DiffNet \cite{wu2019neural} and its extension DiffNet++ \cite{wu2020diffnet++} model the recursive dynamic social diffusion in social recommendation with a layer-wise propagation structure. Wu \emph{et al.} \cite{wu2019dual}  propose a dual graph attention network to collaboratively learn representations for two-fold social effects. Song \emph{et al.} develop DGRec \cite{song2019session} to model both users' session-based interests as well as dynamic social influences. Yu \emph{et al.} \cite{yu2020enhance} propose a deep adversarial framework based on GCNs to address the common issues in social recommendation. In summary, the common idea of these works is to model the user-user and user-item interactions as simple graphs with pairwise connections and then use multiple graph neural layers to capture the node dependencies. 
	
	\subsection{Hypergraph in Recommender Systems}
	Hypergraph \cite{bretto2013hypergraph} provides a natural way to model complex high-order relations and has been extensively employed to tackle various problems. With the development of deep learning, some studies combine GNNs and hypergraphs to enhance representation learning. HGNN \cite{feng2019hypergraph} is the first work that designs a hypergraph convolution operation to handle complex data correlation in representation learning from a spectral perspective. Bai et al. \cite{bai2019hypergraph} introduce hypergraph attention to hypergraph convolutional networks to improve their capacity. However, despite the great capacity in modeling complex data, the potentials of hypergraph for improving recommender systems have been rarely explored. There are only several studies focusing on the combination of these two topics. Bu \emph{et al.} \cite{bu2010music} introduce hypergraph learning to music recommender systems, which is the earliest attempt. The most recent combinations are HyperRec \cite{wang2020next} and DHCF \cite{ji2020dual}, which borrow the strengths of hypergraph neural networks to model the short-term user preference for next-item recommendation and the high-order correlations among users and items for general collaborative filtering, respectively. As for the applications in social recommendation, HMF \cite{zheng2018novel} uses hypergraph topology to describe and analyze the interior relation of social network in recommender systems, but it does not fully exploit high-order social relations since HMF is a hybrid recommendation model.  LBSN2Vec \cite{yang2019revisiting} is a social-aware POI recommendation model that builds hyperedges by jointly sampling friendships and check-ins with random walk, but it focuses on connecting different types of entities instead of exploiting the high-order social network structures. 
	
	\subsection{Self-Supervised Learning}
	Self-supervised learning \cite{hjelm2018learning} is an emerging paradigm to learn with the ground-truth samples obtained from the raw data. It was firstly used in the image domain \cite{bachman2019learning,zhai2019s4l} by rotating, cropping or colorizing the image to create auxiliary supervision signals. The latest advances in this area have extended self-supervised learning to graph representation learning \cite{velickovic2019deep,qiu2020gcc,peng2020graph,sun2019multi}. These studies mainly develop self-supervision tasks from the perspective of investigating graph structure. Node properties such as degree, proximity, and attributes, which are seen as local structure information, are often used as the ground truth to fully exploit the unlabeled data \cite{jin2020self}. For example, InfoMotif \cite{sankar2020beyond} models attribute correlations in motif structures with mutual information maximization to regularize graph neural networks. Meanwhile, global structure information like node pair distance is also harnessed to facilitate representation learning \cite{sun2019multi}. Besides, contrasting congruent and incongruent views of graphs with mutual information maximization \cite{velickovic2019deep,qiu2020gcc} is another way to set up a self-supervised task, which has also shown promising results.
	\par
	As the research of self-supervised learning is still in its infancy, there are only several works combining it with recommender systems \cite{zhou2020s,ma2020disentangled,xin2020self,xia2020self}. These efforts either mine self-supervision signals from future/surrounding sequential data \cite{ma2020disentangled,xin2020self}, or mask attributes of items/users to learn correlations of the raw data \cite{zhou2020s}. However, these thoughts cannot be easily adopted to social recommendation where temporal factors and attributes may not be available. The most relevant work to ours is GroupIM \cite{sankar2020groupim}, which maximizes mutual information between representations of groups and group members to overcome the sparsity problem of group interactions. As the group can be seen as a special social clique, this work can be a corroboration of the effectiveness of social self-supervision signals.
	
	\begin{figure*}[t]
		\centering
		\includegraphics[width=\textwidth]{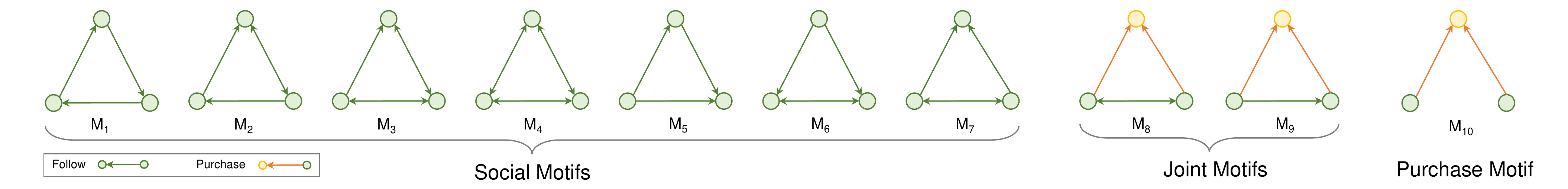}
		\caption{Triangle motifs used in our work. The green circles denote users and the yellow circles denote items. }
		\label{figure.2}
	\end{figure*}

	\section{Proposed Model}
	\subsection{Preliminaries}
	Let $U =\{u_1,u_2, ...,u_m \}$ denote the user set ($|U|=m$), and $I = \{i_1,i_2, ...,i_n\}$ denote the item set ($|I|=n$). $\mathcal{I}(u)$ is the set of user consumption in which items consumed by user $u$ are included. $\bm{R}\in \mathbb{R}^{m\times n}$ is a binary matrix that stores user-item interactions. For each pair $(u,i)$, $r_{ui}=1$ indicates that user $u$ consumed item $i$ while $r_{ui}=0$ means that item $i$ is unexposed to user $u$, or user $u$ is not interested in item $i$. In this paper, we focus on top-K recommendation, and $\hat{r}_{ui}$ denotes the probability of item $i$ to be recommended to user $u$. As for the social relations, we use $\bm{S}\in \mathbb{R}^{m\times m}$ to denote the relation matrix which is asymmetric because we work on directed social networks. In our model, we have multiple convolutional layers, and we use $\{\bm{P}^{(1)}, \bm{P}^{(2)}, \cdots,\bm{P}^{(l)}\} \in \mathbb{R}^{m\times d}$ and $\{\bm{Q}^{(1)}, \bm{Q}^{(2)}, \cdots,\bm{Q}^{(l)}\} \in \mathbb{R}^{n\times d}$ to denote the user and item embeddings of size $d$ learned at each layer, respectively. In this paper, we use bold capital letters to denote matrices and bold lowercase letters to denote vectors.
	\par
	\textbf{Definition 1}: Let $G= (V,E)$ denote a hypergraph, where $V$ is the vertex set containing $N$ unique vertices and $E$ is the edge set containing $M$ hyperedges. Each hyperedge $\epsilon \in E$  can contain any number of vertices and is assigned a positive weight $W_{\epsilon\epsilon}$, and all the weights formulate a diagonal matrix $\bm{W} \in \mathbb{R}^{M \times M}$. The hypergraph can be represented by an incidence matrix $\bm{H} \in \mathbb{R}^{N \times M}$ where $H_{i \epsilon}$ = 1 if the hyperedge $\epsilon \in E$ contains a vertex $v_i \in V$, otherwise 0.  The vertex and edge degree matrices are diagonal matrices denoted by $\bm{D}$ and $\bm{L}$, respectively, where $D_{i i}=\sum_{\epsilon=1}^{M} W_{\epsilon \epsilon} H_{i \epsilon};  L_{\epsilon\epsilon}=\sum_{i=1}^{N} H_{i \epsilon}$. It should be noted that, in this paper, $W_{\epsilon\epsilon}$ is uniformly assigned 1 and hence $\bm{W}$ is an identity matrix.

	\subsection{Multi-Channel Hypergraph Convolutional Network for Social Recommendation}
	In this section, we present our model \textbf{MHCN}, which stands for \textbf{M}ulti-channel \textbf{H}ypergraph \textbf{C}onvolutional \textbf{N}etwork. In Fig. 3, the schematic overview of our model is illustrated.
	\subsubsection{Hypergraph Construction} To formulate the high-order information among users, we first align the social network and user-item interaction graph in social recommender systems and then build hypergraphs over this heterogeneous network. Unlike prior models which construct hyperedges by unifying given types of entities \cite{bu2010music,yang2019revisiting}, our model constructs hyperedges according to the graph structure. As the relations in social networks are often directed, the connectivity of social networks can be of various types. In this paper, we use a set of carefully designed motifs to depict the common types of triangular structures in social networks, which guide the hypergraph construction.
	\par
	Motif, as the specific local structure involving multiple nodes, is first introduced in \cite{milo2002network}. It has been widely used to describe complex structures in a wide range of networks. In this paper, we only focus on triangular motifs because of the ubiquitous triadic closure in social networks, but our model can be seamlessly extended to handle on more complex motifs. Fig. 2 shows all the used triangular motifs. It has been revealed that $\mathcal{M}_{1} - \mathcal{M}_{7}$ are crucial for social computing \cite{benson2016higher}, and we further design $\mathcal{M}_{8} - \mathcal{M}_{10}$ to involve user-item interactions to complement. Given motifs $\mathcal{M}_{1} - \mathcal{M}_{10}$, we categorize them into three groups according to the underlying semantics. $\mathcal{M}_{1} - \mathcal{M}_{7}$ summarize all the possible triangular relations in explicit social networks and describe the high-order social connectivity like `having a mutual friend'. We name this group `\textit{Social Motifs}'. $\mathcal{M}_{8} - \mathcal{M}_{9}$ represent the compound relation, that is, `friends purchasing the same item'. This type of relation can be seen as a signal of strengthened tie, and we name $\mathcal{M}_{8} - \mathcal{M}_{9}$ `\textit{Joint Motifs}'. Finally, we should also consider users who have no explicit social connections. So, $\mathcal{M}_{10}$ is non-closed and defines the implicit high-order social relation that users who are not socially connected but purchased the same item. We name $\mathcal{M}_{10}$ `\textit{Purchase Motif}'. Under the regulation of these three types of motifs, we can construct three hypergraphs that contain different high-order user relation patterns. We use the incidence matrices $\bm{H}^{s}$, $\bm{H}^{j}$ and $\bm{H}^{p}$ to represent these three motif-induced hypergraphs, respectively, where each column of these matrices denotes a hyperedge. For example, in Fig. 3, $\{u_{1},u_{2},u_{3}\}$ is an instance of $\mathcal{M}_{4}$, and we use $e_{1}$ to denote this hyperedge. Then, according to definition 1, we have $H_{u_{1},e_{1}}^{s}=H_{u_{2},e_{1}}^{s}=H_{u_{3},e_{1}}^{s}=1$. 
	
	\begin{figure*}[t]
		\centering
		\includegraphics[width=.95\textwidth]{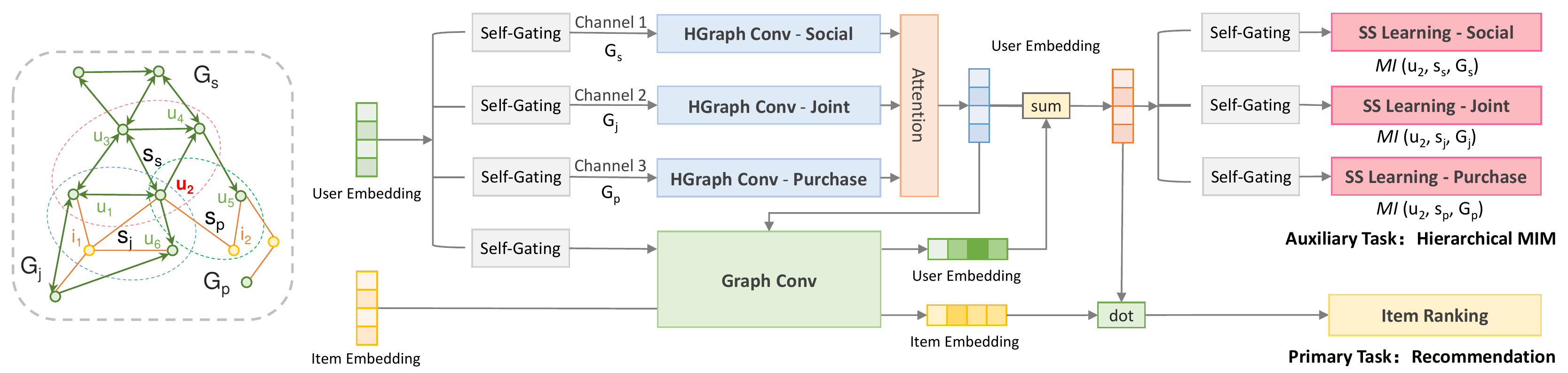}
		\caption{An overview of the proposed model (1-layer). Each triangle in the left graph is a hyperedge and also an instance of defined motifs. $G_{s}$, $G_{j}$ and $G_{p}$ denote the three motif-induced hypergraphs constructed based on social, joint, and purchase motifs, respectively. $s_{s}$, $s_{j}$, and $s_{p}$ in the three dotted ellipses denote three ego-networks with $u_{2}$ as the center, which are subgraphs of $G_{s}$, $G_{j}$ and $G_{p}$, respectively.}
		\label{figure.3}
	\end{figure*}
	
	\subsubsection{Multi-Channel Hypergraph Convolution}
	In this paper, we use a three-channel setting, including `\textit{Social Channel} (s)', `\textit{Joint Channel} (j)', and `\textit{Purchase Channel} (p)', in response to the three types of triangular motifs, but the number of channels can be adjusted to adapt to more sophisticated situations. Each channel is responsible for encoding one type of high-order user relation pattern. As different patterns may show different importances to the final recommendation performance, directly feeding the full base user embeddings $\bm{P}^{(0)}$ to all the channels is unwise. To control the information flow from the base user embeddings $\bm{P}^{(0)}$ to each channel, we design a pre-filter with \emph{self-gating units} (SGUs), which is defined as:
	\begin{equation}
	\bm{P}_{c}^{(0)}=f_{\mathrm{gate}}^{c}(\bm{P}^{(0)})=\bm{P}^{(0)} \odot \sigma(\bm{P}^{(0)}\bm{W}_{g}^{c} +\bm{b}_{g}^{c}),
	\end{equation}
	where $\bm{W}_{g}^{c} \in \mathbb{R}^{d \times d}, \bm{b}_{g}^{c}\in \mathbb{R}^{d}$ are parameters to be learned, $c \in \{s, j, p\}$ represents the channel, $\odot$ denotes the element-wise product and $\sigma$ is the sigmoid nonlinearity. The self-gating mechanism effectively
	serves as a multiplicative skip-connection \cite{dauphin2017language} that learns a nonlinear gate to modulate the base user embeddings at a feature-wise granularity through dimension re-weighting, then we obtain the channel-specific user embeddings $\bm{P}_{c}^{(0)}$.
	\par
	Referring to the spectral hypergraph convolution proposed in \cite{feng2019hypergraph}, we define our hypergraph convolution as:
	\begin{equation}
	\bm{P}^{(l+1)}_{c}=\bm{D}^{-1}_{c} \bm{H}_{c}\bm{L}^{-1}_{c} \bm{H}^{\top}_{c}\bm{P}^{(l)}_{c}.
	\end{equation}
	The difference is that we follow the suggestion in \cite{chen2020revisiting,he2020lightgcn} to remove the learnable matrix for linear transformation and the nonlinear activation function (e.g. leaky ReLU). By replacing $\bm{H}_{c}$ with any of $\bm{H}^{s}$, $\bm{H}^{j}$ and $\bm{H}^{p}$, we can borrow the strengths of hypergraph convolutional networks to learn user representations encoded high-order information in the corresponding channel. As $\bm{D}_{c}$ and $\bm{L}_{c}$ are diagonal matrices which only re-scale embeddings, we skip them in the following discussion. The hypergraph convolution can be viewed as a two-stage refinement performing `node-hyperedge-node' feature transformation upon hypergraph structure. The multiplication operation $\bm{H}_{c}^{\top}\bm{P}^{(l)}_{c}$ defines the message passing from nodes to hyperedges and then premultiplying $\bm{H}_{c}$ is viewed to aggregate information from hyperedges to nodes. However, despite the benefits of hypergraph convolution, there are a huge number of motif-induced hyperedges (e.g. there are 19,385 social triangles in the used dataset, LastFM), which would cause a high cost to build the incidence matrix $\bm{H}_{c}$. But as we only exploit triangular motifs, we show that this problem can be solved in a flexible and efficient way by leveraging the associative property of matrix multiplication. \par
	
	\begin{table}[t]
		\small
		\renewcommand\arraystretch{1.0}
		\caption{Computation of motif-induced adjacency matrices.}
		\label{Table:1}
		\begin{center}
			\begin{tabular}{c|l|l}
				\hline
				Motif&Matrix Computation&$\bm{A}_{M_{i}}=$ \\ \hline
				\hline
				$\mathcal{M}_{1}$ &$\bm{C}=(\bm{U} \bm{U}) \odot \bm{U}^{T}$ & $\bm{C}+\bm{C}^{\top}$ \\
				$\mathcal{M}_{2}$ & $\bm{C}=(\bm{B}  \bm{U}) \odot \bm{U}^{T}+(\bm{U} \bm{B}) \odot \bm{U}^{T}+(\bm{U}  \bm{U}) \odot \bm{B}$ & $\bm{C}+\bm{C}^{\top}$ \\
				$\mathcal{M}_{3}$&$\bm{C}=(\bm{B} \bm{B}) \odot \bm{U}+(\bm{B} \bm{U}) \odot \bm{B}+(\bm{U} \cdot \bm{B}) \odot \bm{B}$ & $\bm{C}+\bm{C}^{\top}$  \\
				$\mathcal{M}_{4}$ &$\bm{C}=(\bm{B} \bm{B}) \odot \bm{B}$ & $\bm{C}$\\
				$\mathcal{M}_{5}$ &$\bm{C}=(\bm{U} \bm{U}) \odot \bm{U}+(\bm{U} \bm{U}^{T}) \odot \bm{U}+(\bm{U}^{T}  \bm{U}) \odot \bm{U}$ & $\bm{C}+\bm{C}^{\top}$ \\
				$\mathcal{M}_{6}$ &$\bm{C}=(\bm{U} \bm{B}) \odot \bm{U}+(\bm{B} \bm{U}^{T}) \odot \bm{U}^{T}+(\bm{U}^{T} \bm{U}) \odot \bm{B}$ & $\bm{C}$\\
				$\mathcal{M}_{7}$ &$\bm{C}=(\bm{U}^{T} \bm{B}) \odot \bm{U}^{T}+(\bm{B} \bm{U}) \odot \bm{U}+(\bm{U}  \bm{U}^{T}) \odot \bm{B}$ & $\bm{C}$\\
				$\mathcal{M}_{8}$ &$\bm{C}=(\bm{R} \bm{R}^{T}) \odot \bm{B}$ & $\bm{C}$\\
				$\mathcal{M}_{9}$ &$\bm{C}=(\bm{R} \bm{R}^{T}) \odot \bm{U}$& $\bm{C}+\bm{C}^{\top}$\\
				$\mathcal{M}_{10}$ &$\bm{C}=\bm{R} \bm{R}^{T} $ & $\bm{C}$\\
				\hline	
			\end{tabular}
		\end{center}
	\end{table}
	
	Following \cite{zhao2018ranking}, we let $\bm{B}=\bm{S}\odot\bm{S}^{T}$ and $\bm{U}=\bm{S}-\bm{B}$ be the adjacency matrices of the bidirectional and unidirectional social networks respectively. We use $\bm{A}_{M_{k}}$ to represent the motif-induced adjacency matrix and $(\bm{A}_{M_{k}})_{i,j}=1$ means that vertex $i$ and vertex $j$ appear in one instance of $\mathcal{M}_{k}$. As two vertices can appear in multiple instances of $\mathcal{M}_{k}$, $(\bm{A}_{M_{k}})_{i,j}$ is computed by:
	\begin{equation}
	(\bm{A}_{M_{k}})_{i,j}=\#(i, j \text{ occur in the same instance of } \mathcal{M}_{k}).
	\end{equation}
	Table 1 shows how to calculate $\bm{A}_{M_{k}}$ in the form of matrix multiplication. As all the involved matrices in Table 1 are sparse matrices, $\bm{A}_{M_{k}}$ can be efficiently calculated.  Specifically, the basic unit in Table 1 is in a general form of $\bm{XY} \odot \bm{Z}$, which means $\bm{A}_{M_{1}}$ to $\bm{A}_{M_{9}}$ may be sparser than $\bm{Z}$ (i.e. $\bm{B}$ or $\bm{U}$) or as sparse as $\bm{Z}$. $\bm{A}_{M_{10}}$ could be a little denser, but we can filter out the popular items (we think consuming popular items might not reflect the users' personalized preferences) when calculating $\bm{A}_{M_{10}}$ and remove the entries less than a threshold (e.g. 5) in $\bm{A}_{M_{10}}$ to keep efficient calculation. For symmetric motifs, $\bm{A}_{M}=\bm{C}$, and for the asymmetric ones $\bm{A}_{M}=\bm{C}+\bm{C}^{T}$. Obviously, without considering self-connection, the summation of $\bm{A}_{M_{1}}$ to $\bm{A}_{M_{7}}$ is equal to $\bm{H}^{s}\bm{H}^{s\top}$, as each entry of $\bm{H}^{s}\bm{H}^{s\top}\in \mathbb{R}^{m\times m}$ also indicates how many social triangles contain the node pair represented by the row and column index of the entry. Analogously, the summation of $\bm{A}_{M_{8}}$ to $\bm{A}_{M_{9}}$ is equal to $\bm{H}^{j}\bm{H}^{j\top}$ without self-connection and $\bm{A}_{M_{10}}$ is equal to $\bm{H}^{p}\bm{H}^{p\top}$. Taking the calculation of $\bm{A}_{M_{1}}$ as an example, it is evident that $\bm{U} \bm{U}$ constructs a unidirectional path connecting three vertices, and the operation $\odot \bm{U}$ makes the path into a loop, which is an instance of $\bm{A}_{M_{1}}$.  As $\bm{A}_{M_{10}}$ also contains the triangles in $\bm{A}_{M_{8}}$ and $\bm{A}_{M_{9}}$. So, we remove the redundancy from $\bm{A}_{M_{10}}$. Finally, we use $\bm{A}_{s} = \sum_{k=1}^{7}\bm{A}_{M_{k}}$, $\bm{A}_{j} = \bm{A}_{M_{8}}+\bm{A}_{M_{9}}$, and $\bm{A}_{p} = \bm{A}_{M_{10}}-\bm{A}_{j}$ to replace $\bm{H}^{s}\bm{H}^{s\top}$, $\bm{H}^{j}\bm{H}^{j\top}$, and $\bm{H}^{p}\bm{H}^{p\top}$ in Eq. (2), respectively. Then we have a transformed hypergraph convolution, defined as:
	\begin{equation}
	\bm{P}^{(l+1)}_{c}=\hat{\bm{D}}^{-1}_{c} \bm{A}_{c}\bm{P}^{(l)}_{c},
	\end{equation}
	where $\hat{\bm{D}}_{c}\in \mathbb{R}^{m\times m}$ is the degree matrix of $\bm{A}_{c}$. Obviously, Eq (4) is equivalent to Eq (2), and can be a simplified substitution of the hypergraph convolution. Since we follow the design of LightGCN which has subsumed the effect of self-connection, and thus skipping self-connection in adjacency matrix does not matter too much. In this way, we circumvent the individual hyperedge construction and computation, and greatly reduce the computational cost.
	\par
	\subsubsection{Learning Comprehensive User Representations}
	After propagating the user embeddings through $L$ layers, we average the embeddings obtained at each layer to form the final channel-specific user representation: $\bm{P}^{*}_{c}=\frac{1}{L+1}\sum_{l=0}^{L}\bm{P}^{(l)}_{c}$ to avoid the over-smoothing problem \cite{he2020lightgcn}. Then we use the attention mechanism \cite{vaswani2017attention} to selectively aggregate information from different channel-specific user embeddings to form the comprehensive user embeddings. For each user $u$, a triplet $(\alpha^{s},\alpha^{j},\alpha^{p})$ is learned to measure the different contributions of the three channel-specific embeddings to the final recommendation performance. The attention function $f_{\mathrm{att}}$ is defined as:
	\begin{equation}
	\alpha_{c}=f_{\mathrm{att}}({p}^{*}_{c})=\frac{\exp (\bm{a}^{\top} \cdot\bm{W}_{att} \bm{p}^{*}_{c})}{\sum_{c^{\prime} \in \{s,j,p\}} \exp (\bm{a}^{\top} \cdot\bm{W}_{att} \bm{p}^{*}_{c^{\prime}})},
	\end{equation}
	where $\bm{a}\in\mathbb{R}^{d}$ and $\bm{W}_{att}\in\mathbb{R}^{d\times d}$ are trainable parameters, and the comprehensive user representation $\bm{p}^{*}=\sum_{c \in \{s,j,p\}} \alpha_{c}\bm{p}^{*}_{c},$. \par
	Note that, since the explicit social relations are noisy and isolated relations are not a strong signal of close friendship \cite{yu2018adaptive,yu2017hybrid}, we discard those relations which are not part of any instance of defined motifs. So, we do not have a convolution operation directly working on the explicit social network $\bm{S}$. Besides, in our setting, the hypergraph convolution cannot directly aggregate information from the items (we do not incorporate the items into $\bm{A}_{j}$ and $\bm{A}_{p}$). To tackle this problem, we additionally perform simple graph convolution on the user-item interaction graph to encode the purchase information and complement the multi-channel hypergraph convolution. The simple graph convolution is defined as:
	\begin{equation}
	\begin{split}
	&\bm{P}^{(l+1)}_{r}=\bm{D}^{-1}_{u} \bm{R}\bm{Q}^{(l)},\ \ \bm{P}^{(0)}_{r}=f_{\mathrm{gate}}^{r}(\bm{P}^{(0)}),\\
	&\bm{Q}^{(l+1)}=\bm{D}^{-1}_{i} \bm{R}^{\top}\bm{P}^{(l)}_{m}, \ \ \bm{P}^{(l)}_{m}=\sum_{c \in \{s,j,p\}} \alpha_{c}\bm{p}^{(l)}_{c}+\frac{1}{2}\bm{P}^{(l)}_{r},
	\end{split}
	\end{equation}
	where $\bm{P}^{(l)}_{r}$ is the gated user embeddings for simple graph convolution, $\bm{P}^{(l)}_{m}$ is the combination of the comprehensive user embeddings and $\bm{P}^{(l)}_{r}$, and $\bm{D}_{u} \in \mathbb{R}^{m\times m}$ and $\bm{D}_{i} \in \mathbb{R}^{n\times n}$ are degree matrices of $\bm{R}$ and $\bm{R}^{\top}$, respectively. Finally, we obtain the final user and item embeddings $\bm{P}$ and $\bm{Q}$ defined as:
	\begin{equation}
	\bm{P} = \bm{P}^{*}+\frac{1}{L+1}\sum_{l=0}^{L}\bm{P}^{(l)}_{r}, \ \ \bm{Q} = \frac{1}{L+1}\sum_{l=0}^{L}\bm{Q}^{(l)},
	\end{equation}
	where $\bm{P}^{(0)}$ and $\bm{Q}^{(0)}$ are randomly initialized.
	\par

	\subsubsection{Model Optimization}
	To learn the parameters of MHCN, we employ the Bayesian Personalized Ranking (BPR) loss \cite{rendle2009bpr}, which is a pairwise loss that promotes an observed entry to be ranked higher than its unobserved counterparts:
	\begin{equation}
	\mathcal{L}_{r}=\sum_{i \in \mathcal{I}(u), j \notin \mathcal{I}(u)}-\log \sigma\left(\hat{r}_{u,i}(\Phi)-\hat{r}_{u,j}(\Phi)\right)+\lambda\|\Phi\|_{2}^{2},
	\end{equation}
	where $\Phi$ denotes the parameters of MHCN, $\hat{r}_{u,i}=\bm{p}_{u}^{\top}\bm{q}_{i}$ is the predicted score of $u$ on $i$, and $\sigma(\cdot)$ here is the sigmoid function. Each time a triplet including the current user $u$, the positive item $i$ purchased by $u$, and the randomly sampled negative item $j$ which is disliked by $u$ or unknown to $u$, is fed to MHCN. The model is optimized towards ranking $i$ higher than $j$ in the recommendation list for $u$. In addition, $L_{2}$ regularization with the hyper-parameter $\lambda$ is imposed to reduce generalized errors.

	\subsection{Enhancing MHCN with Self-Supervised Learning}
	Owing to the exploitation of high-order relations, MHCN shows great performance (reported in Table 3 and 4). However, a shortcoming of MHCN is that the aggregation operations (Eq. 5 and 6) might lead to a loss of high-order information, as different channels would learn embeddings with varying distributions on different hypergraphs \cite{yu2014mixed}. Concatenating the embeddings from different channels could be the alternative, but it uniformly weighs the contributions of different types of high-order information in recommendation generation, which is not in line with the reality and leads to inferior performance in our trials. To address this issue and fully inherit the rich information in the hypergraphs, we innovatively integrate self-supervised learning into the training of MHCN. \par
	
	In the scenarios of representation learning, self-supervised task usually either serves as a pretraining strategy or an auxiliary task to improve the primary task \cite{jin2020self}. In this paper, we follow the \emph{primary \& auxiliary} paradigm, and set up a self-supervised auxiliary task to enhance the recommendation task (primary task). The recent work \textit{Deep Graph Infomax} (DGI) \cite{velickovic2019deep} is a general and popular approach for learning node representations within graph-structured data in a self-supervised manner. It relies on maximizing mutual information (MI) between node representations and corresponding high-level summaries of graphs. However, we consider that the graph-node MI maximization stays at a coarse level and there is no guarantee that the encoder in DGI can distill sufficient information from the input data. Therefore, with the increase of the graph scale, the benefits brought by MI maximization might diminish.  For a better learning method which fits our scenario more, we inherit the merits of DGI to consider mutual information and further extend the graph-node MI maximization to a fine-grained level by exploiting the hierarchical structure in hypergraphs.
	\par
	
		\begin{figure}[t]
		\centering
		\includegraphics[width=0.45\textwidth]{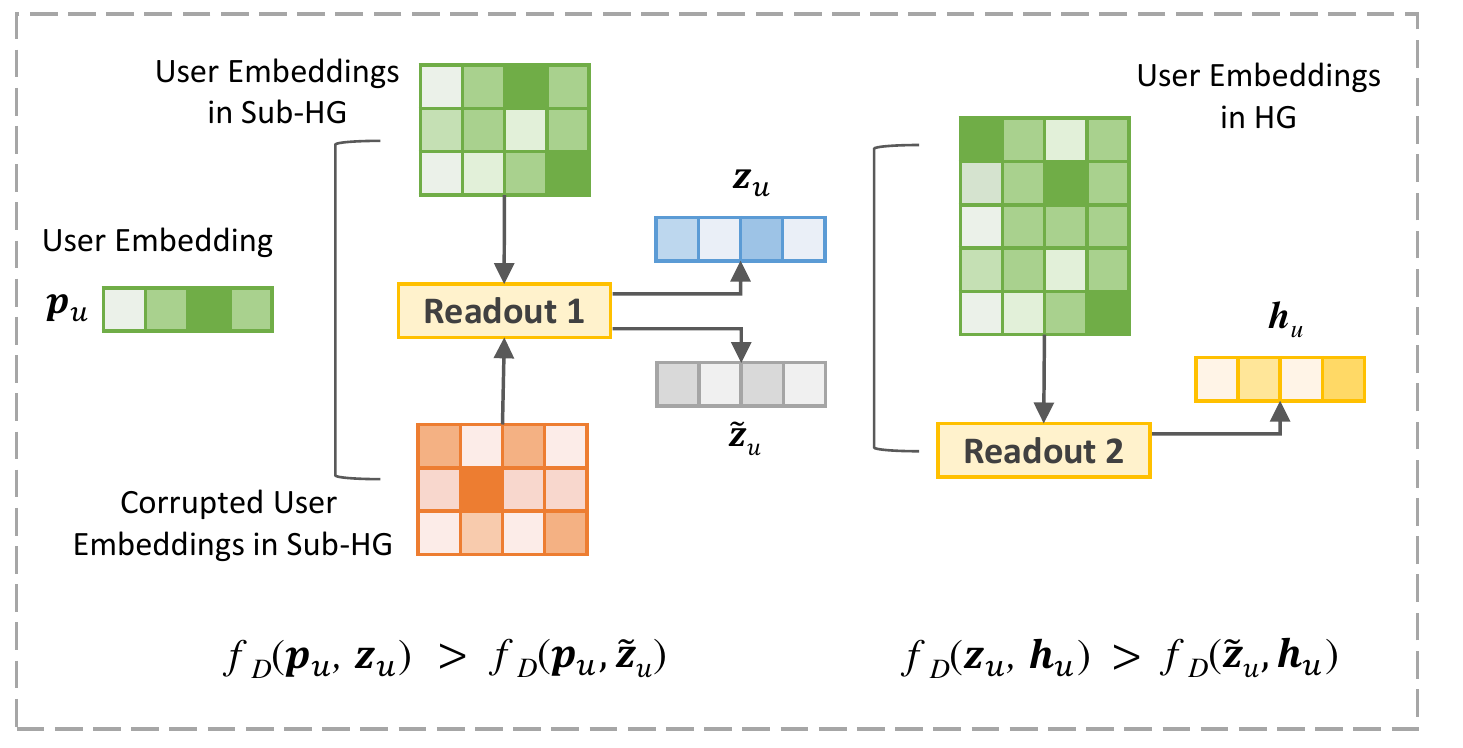}
		\caption{Hierarchical mutual information maximization on hypergraphs.}
		\label{figure.4}
		\vspace{-15px}
	\end{figure}

	Recall that, for each channel of MHCN, we build the adjacency matrix $\bm{A}_{c}$ to capture the high-order connectivity information. Each row in $\bm{A}_{c}$ represents a subgraph of the corresponding hypergraph centering around the user denoted by the row index. Then we can induce a hierarchy: `user node $\leftarrow$ user-centered sub-hypergraph $\leftarrow$ hypergraph' and create self-supervision signals from this structure. Our intuition of the self-supervised task is that the comprehensive user representation should reflect the user node's local and global high-order connectivity patterns in different hypergraphs, and this goal can be achieved by hierarchically maximizing the mutual information between representations of the user, the user-centered sub-hypergraph, and the hypergraph in each channel. The mutual information measures the structural informativeness of the sub- and the whole hypergraph towards inferring the user preference through the reduction in local and global structure uncertainty. \par
	To get the sub-hypergraph representation, instead of averaging the embeddings of the users in the sub-hypergraph, we design a readout function $f_{\mathrm{out_{1}}}: \mathbb{R}^{k\times d}\rightarrow\mathbb{R}^{d}$, which is permutation-invariant and formulated as:
	\begin{equation}
	\bm{z}^{c}_{u}=f_{out_{1}}(\bm{P}_{c},\bm{a}^{c}_{u})=\frac{\bm{P}_{c}\bm{a}^{c}_{u}}{sum(\bm{a}^{c}_{u})},
	\end{equation}
	where $\bm{P}_{c}=f^{c}_{\mathrm{gate}}(\bm{P})$ is to control the participated magnitude of $\bm{P}$ to avoid overfitting and mitigate gradient conflict between the primary and auxiliary tasks, $\bm{a}^{c}_{u}$ is the row vector of $\bm{A}_{c}$ corresponding to the center user $u$, and $sum(\bm{a}^{c}_{u})$ denotes how many connections in the sub-hypergraph. In this way, the weight (importance) of each user in the sub-hypergraph is considered to form the sub-hypergraph embedding $\bm{z}_{u}$. Analogously, we define the other readout function $f_{\mathrm{out_{2}}}: \mathbb{R}^{m\times d}\rightarrow\mathbb{R}^{d}$, which is actually an average pooling to summarize the obtained sub-hypergraph embeddings into a graph-level representation: 
	\begin{equation}
	\bm{h}^{c}=f_{out_{2}}(\bm{Z}_{c})=\mathrm{AveragePooling}(\bm{Z}_{c}).
	\end{equation}
	\par
	We tried to use InfoNCE \cite{oord2018representation} as our learning objective to maximize the hierarchical mutual information. But we find that the pairwise ranking loss, which has also been proved to be effective in mutual information estimation \cite{kemertas2020rankmi}, is more compatible with the recommendation task. We then define the objective function of the self-supervised task as follows:
	\begin{equation}
	\begin{split}
	\mathcal{L}_{s}=-\sum_{c \in \{s,j,p\}} \Big \{\sum_{u \in U}\log \sigma(f_{D}(\bm{p}^{c}_{u},\bm{z}^{c}_{u})-f_{D}(\bm{p}^{c}_{u},\bm{\tilde{z}}^{c}_{u}))\\
	+\sum_{u \in U}\log \sigma(f_{D}(\bm{z}^{c}_{u},\bm{h}^{c})-f_{D}(\bm{\tilde{z}}^{c}_{u},\bm{h}^{c}))\Big \}.
	\end{split}	
	\end{equation}
	$f_{D}(\cdot): \mathbb{R}^{d} \times \mathbb{R}^{d} \longmapsto \mathbb{R}$ is the discriminator function that takes two vectors as the input and then scores the agreement between them. We simply implement the discriminator as the dot product between two representations. Since there is a bijective mapping between $\bm{P}_{c}$ and $\bm{Z}_{c}$, they can be the ground truth of each other. We corrupt $\bm{Z_{c}}$  by both row-wise and column-wise shuffling to create negative examples $\bm{\tilde{Z}}_{c}$. We consider that, the user should have a stronger connection with the sub-hypergraph centered with her  (local structure), so we directly maximize the mutual information between their representations. By contrast, the user would not care all the other users too much (global structure), so we indirectly maximize the mutual information between the representations of the user and the complete hypergraph by regarding the sub-hypergraph as the mediator. Compared with DGI which only maximizes the mutual information between node and graph representations, our hierarchical design can preserve more structural information of the hypergraph into the user representations (comparison is shown in Section 4.3). Figure 4 illustrates the hierarchical mutual information maximization. 
	\par
	Finally, we unify the objectives of the recommendation task (primary) and the task of maximizing hierarchical mutual information (auxiliary) for joint learning. The overall objective is defined as:
	\begin{equation}
	\mathcal{L}=\mathcal{L}_{r}+\beta\mathcal{L}_{s},
	\end{equation}
	where $\beta$ is a hyper-parameter used to control the effect of the auxiliary task and $\mathcal{L}_{s}$ can be seen as a regularizer leveraging hierarchical structural information of the hypergraphs to enrich the user representations in the recommendation task for a better performance.
	
	\subsection{Complexity Analysis}
	In this section, we discuss the complexity of our model.\par 
	\textbf{Model size}. The trainable parameters of our model consist of three parts: user and item embeddings, gate parameters, and attention parameters. For the first term, we only need to learn the $0^{th}$ layer user embeddings $\bm{P}^{(0)} \in \mathbb{R}^{m\times d}$ and item embeddings $\bm{Q}^{(0)} \in \mathbb{R}^{n\times d}$. As for the second term, we employ seven gates, four for MHCN and three for the self-supervised task. Each of the gate has parameters of size $(d+1)\times d$, while the attention parameters are of the same size. To sum up, the model size approximates $(m+n+8d)d$ in total. As $\min(m,n)\gg d$, our model is fairly light.\par
	\textbf{Time complexity}. 
	The computational cost mainly derives from four parts: hypergraph/graph convolution, attention, self-gating, and mutual information maximization. For the multi-channel hypergraph convolution through $L$ layers, the propagation consumption is less than $\mathcal{O}(|\bm{A}^{+}|dL)$, where $|\bm{A}^{+}|$ denotes the number of nonzero elements in $\bm{A}$, and here $|\bm{A}^{+}|=\max(|\bm{A}^{+}_{s}|,|\bm{A}^{+}_{j}|,|\bm{A}^{+}_{p}|)$. Analogously, the time complexity of the graph convolution is $\mathcal{O}(|\bm{R}^{+}|dL)$. As for the attention and self gating mechanism, they both contribute $\mathcal{O}(md^{2})$ time complexity. The cost of mutual information maximization is mainly from $f_{out_{1}}$, which is $\mathcal{O}(|\bm{A}^{+}|d)$. Since we follow the setting in \cite{he2020lightgcn} to remove the learnable matrix for linear transformation and the nonlinear activation function, the time complexity of our model is much lower than that of previous GNNs-based social recommendation models.

	\begin{table}[!htb]
		\renewcommand\arraystretch{1.0}
		\caption{Dataset Statistics}

		\label{Table:2}
		\begin{center}
			\begin{tabular}{c|ccccc}
				\hline
				Dataset&\#User & \#Item &  \#Feedback &  \#Relation & Density\\ \hline
				\hline
				LastFM &1,892 &  17,632 & 92,834 & 25,434  & 0.28\%\\
				Douban & 2,848 & 39,586 & 894,887 &35,770 & 0.79\%\\
				Yelp&19,539  &21,266 & 450,884 &864,157 & 0.11\%\\
				\hline
			\end{tabular}
		\end{center}
	\end{table}

\begin{table*}[t]
\centering
\caption{General recommendation performance comparison.}
\label{Table:3}
\renewcommand\arraystretch{1.0}
\begin{center}
\begin{tabular}{cc|cccccccc|c|c}
\hline
Dataset&Metric&GraphRec&BPR&SBPR&DiffNet++&DHCF&LightGCN&\textbf{MHCN}&$\bm{S}^{2}$\textbf{-MHCN}&Improv.&$\bm{S}^{2}$-Improv.\\ \hline
\hline
\multirow{3}{*}{LastFM}
&P@10&17.385\%&16.556\%&16.893\%&18.485\%&16.877\%&19.205\%&19.625\%&\textbf{20.052}\%&4.410\%&2.175\%\\		
&R@10&18.020\%&16.803\%&17.245\%&18.737\%&17.131\%&19.480\%&19.945\%&\textbf{20.375}\%&4.594\%&2.155\%\\			
&N@10&0.21173&0.19943&0.20516&0.22310&0.20744&0.23392&0.23834&\textbf{0.24395}&4.287\%&2.156\%\\

\hline
\multirow{3}{*}{Douban}
&P@10&17.021\%&15.673\%&15.993\%&17.532\%&16.871\%&17.780\%&18.283\%&\textbf{18.506}\%&4.083\%&1.220\%\\		
&R@10&5.916\%&5.160\%&5.322\%&6.205\%&5.755\%&6.247\%&6.556\%&\textbf{6.681}\%&6.947\%&1.906\%\\		
&N@10&0.19051&0.17476&0.17821&0.19701&0.18655&0.19881&0.20694&\textbf{0.21038}&5.819\%&1.662\%\\			

\hline
\multirow{3}{*}{Yelp}
&P@10&2.323\%&2.002\%&2.192\%&2.480\%&2.298\%&2.586\%&2.751\%&\textbf{3.003}\%&16.125\%&9.160\%\\			
&R@10&6.075\%&5.173\%&5.468\%&6.354\%&5.986\%&6.525\%&6.862\%&\textbf{7.885}\%&17.247\%&14.908\%\\			
&N@10&0.04653&0.03840&0.04314&0.04833&0.04700&0.04998&0.05356&\textbf{0.06061}&21.268\%&13.162\%\\

\hline
\end{tabular}
\end{center}
\end{table*}

\begin{table*}[t]
\centering	
\caption{Cold-start recommendation performance comparison.}
\label{Table:4}
\renewcommand\arraystretch{1.1}
\begin{center}
\begin{tabular}{cc|cccccccc|c|c}
\hline
Dataset&Metric&GraphRec&BPR&SBPR&DiffNet++&DHCF&LightGCN&\textbf{MHCN}&$\bm{S}^{2}$\textbf{-MHCN}&Improv.& $\bm{S}^{2}$-Improv.\\ \hline
\hline
\multirow{3}{*}{LastFM}
&P@10&4.662\%&3.784\%&4.573\%&5.102\%&3.974\%&4.809\%&5.466\%&\textbf{5.759}\%&12.877\%&5.360\%\\		
&R@10&18.033\%&15.240\%&18.417\%&21.365\%&16.395\%&20.361\%&23.354\%&\textbf{24.431}\%&14.350\%&4.611\%\\			
&N@10&0.14675&0.12460&0.15141&0.16031&0.14285&0.15044&0.17218&\textbf{0.19138}&19.381\%&11.151\%\\

\hline
\multirow{3}{*}{Douban}
&P@10&2.007\%&1.722\%&1.935\%&2.230\%&1.921\%&2.134\%&2.343\%&\textbf{2.393}\%&7.309\%&2.133\%\\		
&R@10&8.215\%&7.178\%&8.084\%&8.705\%&7.977\%&8.317\%&9.646\%&\textbf{10.632}\%&22.136\%&10.227\%\\		
&N@10&0.05887&0.04784&0.05716&0.06767&0.05533&0.06037&0.06771&\textbf{0.07113}&5.113\%&5.052\%	\\	

\hline
\multirow{3}{*}{Yelp}
&P@10&1.355\%&1.232\%&1.286\%&1.475\%&1.314\%&1.504\%&1.545\%&\textbf{1.747}\%&14.108\%&13.074\%\\			
&R@10&5.901\%&5.468\%&5.720\%&6.635\%&5.876\%&6.753\%&6.838\%&\textbf{7.881}\%&12.264\%&15.253\%\\			
&N@10&0.03896&0.03448&0.03671&0.04237&0.03826&0.04273&0.04354&\textbf{0.05143}&15.703\%&18.121\%\\

\hline
\end{tabular}
\end{center}
\end{table*}

	\section{Experiments and Results}
	\subsection{Experimental Settings} 
	\noindent\textbf{Datasets.} Three real-world datasets: LastFM\footnote{http://files.grouplens.org/datasets/hetrec2011/}, Douban\footnote{https://pan.baidu.com/s/1hrJP6rq}, and Yelp \cite{yin2019social} are used in our experiments. As our aim is to generate Top-K recommendation, for Douban which is based on explicit ratings, we leave out ratings less than 4 and assign 1 to the rest. The statistics of the datasets is shown in Table 2. We perform 5-fold cross-validation on the three datasets and report the average results. \par
	\noindent\textbf{Baselines.} We compare MHCN with a set of strong and commonly-used baselines including MF-based and GNN-based models:
	\begin{itemize}[leftmargin=*]
		\item \textbf{BPR} \cite{rendle2009bpr} is a popular recommendation model based on Bayesian personalized ranking. It models the order of candidate items by a pairwise ranking loss.
		\item \textbf{SBPR} \cite{zhao2014leveraging} is a MF based social recommendation model which extends BPR and leverages social connections to model the relative order of candidate items.
		\item \textbf{LightGCN} \cite{he2020lightgcn} is an efficient GCN-based general recommendation model that leverages the user-item proximity to learn node representations and generate recommendations.
	    \item \textbf{GraphRec} \cite{fan2019graph} is the first GNN-based social recommendation model that models both user-item and user-user interactions.
		\item \textbf{DiffNet++} \cite{wu2020diffnet++} is the latest GCN-based social recommendation method that models the recursive dynamic social diffusion in both the user and item spaces.
		\item \textbf{DHCF} \cite{ji2020dual} is a recent hypergraph convolutional network-based method that models the high-order correlations among users and items for general recommendation.
	\end{itemize} 
	Two versions of the proposed multi-channel hypergraph convolutional network are investigated in the experiments. \textbf{MHCN} denotes the vanilla version and $\bm{S}^{2}$\textbf{-MHCN} denotes the self-supervised version.\par
	\noindent\textbf{Metrics.} To evaluate the performance of all methods, two relevancy-based metrics \emph{Precision@10} and \emph{Recall@10} and one ranking-based metric \emph{NDCG@10} are used. We perform item ranking on all the candidate items instead of the sampled item sets to calculate the values of these three metrics, which guarantees that the evaluation process is unbiased.\\
	\noindent\textbf{Settings.} For a fair comparison, we refer to the best parameter settings reported in the original papers of the baselines and then use grid search to fine tune all the hyperparameters of the baselines to ensure the best performance of them. For the general settings of all the models, the dimension of latent factors (embeddings) is empirically set to 50, the regularization coefficient $\lambda=0.001$, and the batch size is set to 2000. We use Adam to optimize all these models. Section 4.4 reports the influence of different parameters (i.e. $\beta$ and the depth) of MHCN, and we use the best parameter settings in Section 4.2, and 4.3. 
	\par

	\subsection{Recommendation Performance}
	In this part, we validate if MHCN outperforms existing social recommendation baselines. Since the primary goal of social recommendation is to mitigate data sparsity issue and improve the recommendation performance for cold-start users. Therefore, we respectively conduct experiments on the complete test set and the cold-start test set in which only the cold-start users with less than 20 interactions are contained. The experimental results are shown in Table 3 and Table 4. The improvement is calculated by subtracting the best performance value of the baselines from that of $S^{2}$-MHCN and then using the difference to divide the former. Analogously, $S^{2}$-improvement is calculated by comparing the values of the performance of MHCN and and $\bm{S}^{2}$-MHCN. According to the results, we can draw the following conclusions:

	\begin{itemize}[leftmargin=*]
	\item MHCN shows great performance in both the general and cold-start recommendation tasks. Even without self-supervised learning, it beats all the baselines by a fair margin. Meanwhile, 	self-supervised learning has great ability to further improve MHCN. Compared with the vanilla version, the self-supervised version shows decent improvements in all the cases. Particularly, in the cold-start recommendation task, self-supervised learning brings significant gains. On average, $\bm{S}^{2}$-MHCN achieves about 5.389\% improvement in the general recommendation task and 9.442\% improvement in the cold-start recommendation task compared with MHCN. Besides, it seems that, the sparser the dataset, the more improvements self-supervised learning brings.
	\item GNN-based recommendation models significantly outperform the MF-based recommendation models. Even the general recommendation models based on GNNs show much better performance than MF-based social recommendation models. However, when compared with the counterparts based on the same building block (i.e. MF-based \emph{vs.} MF-based, GNNs-based  \emph{vs.} GNNs-based), social recommendation models are still competitive and by and large outperform the corresponding general recommendation models except LightGCN. 
	\item LightGCN is a very strong baseline. Without considering the two variants of MHCN, LightGCN shows the best or the second best performance in most cases. This can be owed to the removal of the redundant operations including the nonlinear activation function and transformation matrices.  The other baselines such as GraphRec might be limited by these useless operations, and fail to outperform LightGCN, though the social information is incorporated.
	\item Although DHCF is also based on hypergraph convolution, it does not show any competence in all the cases. We are unable to reproduce its superiority reported in the original paper \cite{ji2020dual}. There are two possible causes which might lead to its failure. Firstly, it only exploits the user-item high-order relations. Secondly, the way to construct hyperedges is very impractical in this model, which leads to a very dense incidence matrix. The model would then encounter the over-smoothing problem and suffer from heavy computation. 
	\end{itemize}

\begin{figure}[t]
	\centering
	\includegraphics[width=0.5\textwidth]{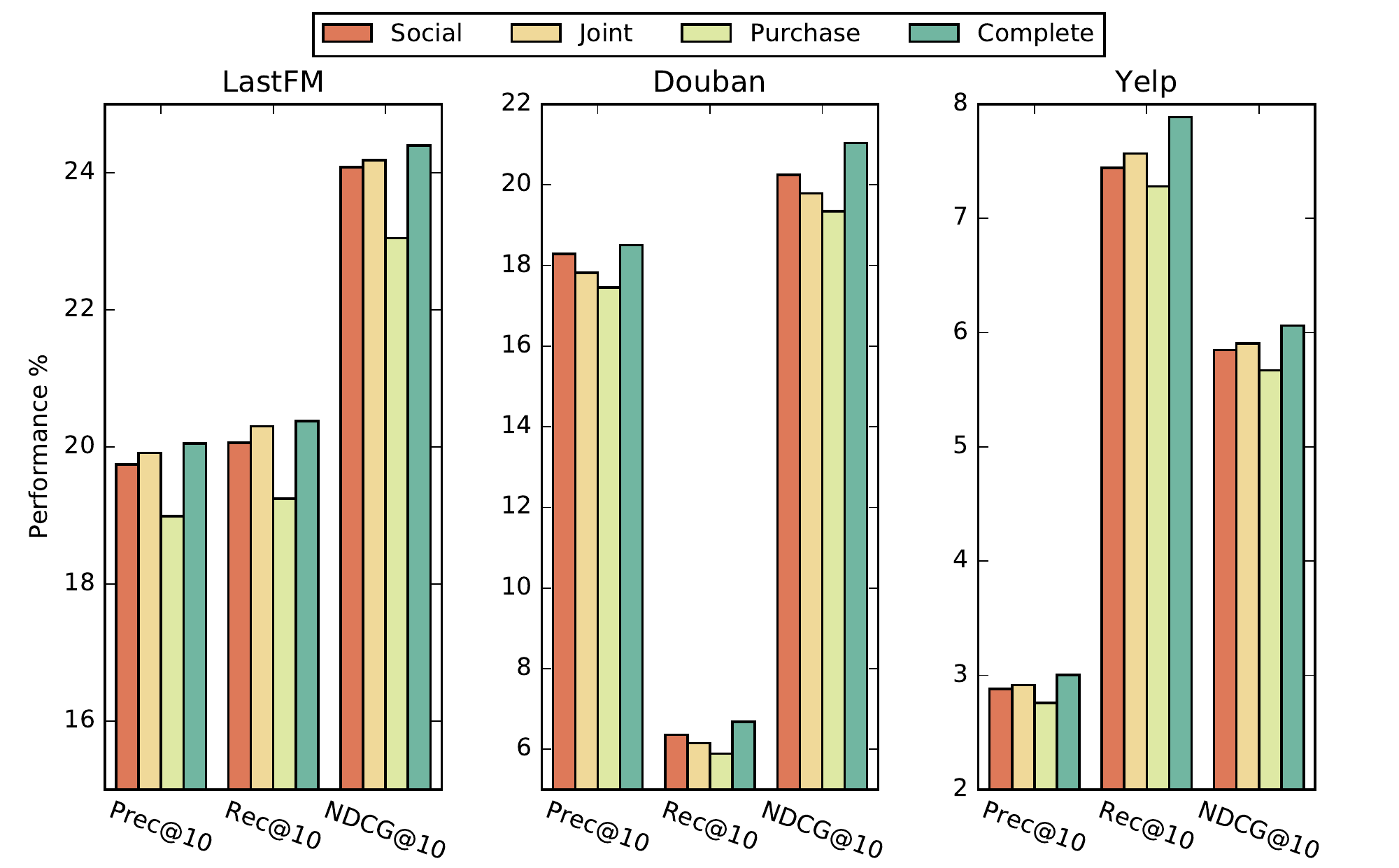}
	\caption{Contributions of each channel on different datasets.}
	\label{figure.5}
	\vspace{-10px}
\end{figure}

\begin{figure}[t]
	\centering
	\includegraphics[width=0.5\textwidth]{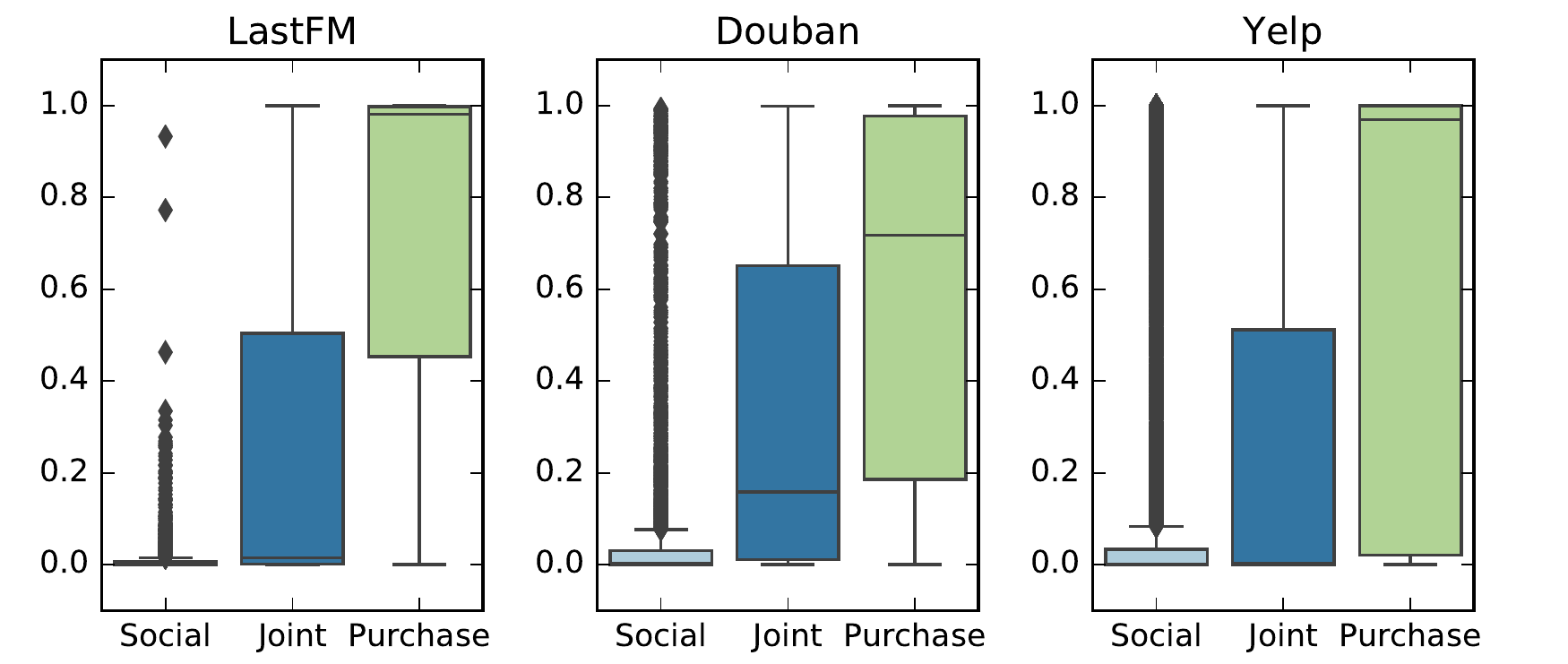}
	\caption{The distributions of the attention weights on different datasets.}
	\label{figure.6}
	\vspace{-10px}
\end{figure}
	\subsection{Ablation Study}
    In this section, we conduct an ablation study to investigate the interplay of the components in $S^{2}$-MHCN and validate if each component positively contributes to the final recommendation performance. 
	\subsubsection{Investigation of Multi-Channel Setting}
	We first investigate the multi-channel setting by removing any of the three channels from $S^{2}$-MHCN and leaving the other two to observe the changes of performance. Each bar in the plots (except \emph{complete}) represents the case that the corresponding channel is removed, while \emph{complete} means no module has been removed. From Fig. 5, we can observe that removing any channel would cause performance degradation. But it is obvious that \emph{purchase channel} contributes the most to the final performance. Without this channel, $S^{2}$-MHCN falls to the level of LightGCN shown in Table 3. By contrast, removing \emph{Social channel} or \emph{Joint channel} would not have such a large impact on the final performance. Comparing \emph{Social channel} with \emph{Joint channel}, we can observe that the former contributes slightly more on LastFM and Yelp, while the latter, in terms of the performance contribution, is more important on Douban. \par
	To further investigate the contribution of each channel when they are all employed, we visualize the attention scores learned along with other model parameters, and draw a box plot to display the distributions of the attention weights. According to Fig. 6, we can observe that, for the large majority of users in LastFM, \emph{Social channel} has limited influence on the comprehensive user representations. In line with the conclusions from Fig. 5, \emph{Purchase channel} plays the most important role in shaping the comprehensive user representations. The importance of \emph{Joint channel} falls between the other two. The possible reason could be that, social relations are usually noisy and the users who are only socially connected might not always share similar preferences. 
	
		\begin{figure}[t]
		\centering
		\includegraphics[width=0.5\textwidth]{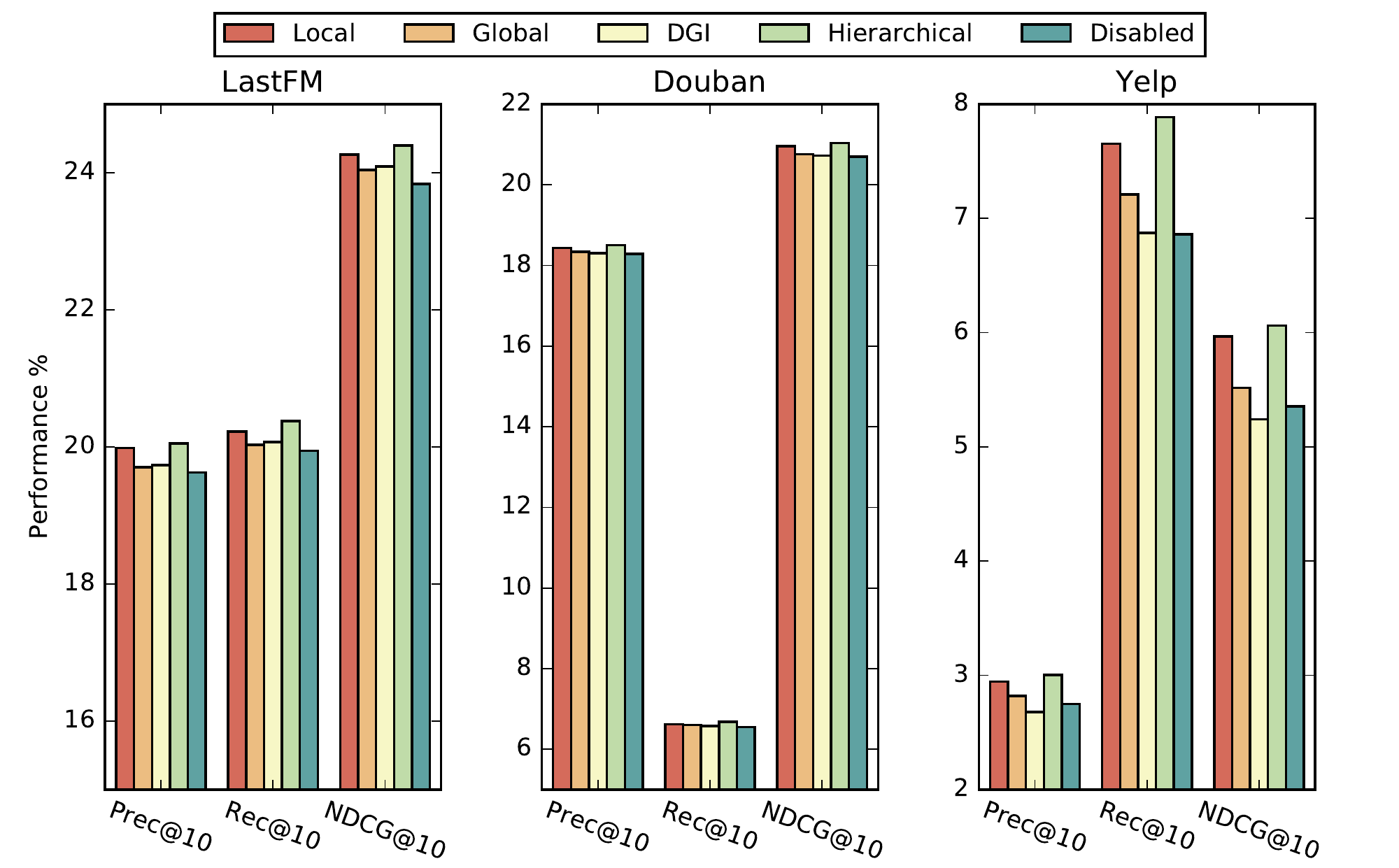}
		\caption{Investigation of Hierarchical Mutual Information Maximization on different datasets.}
		\label{figure.7}
		\vspace{-10px}
	\end{figure}

	\subsubsection{Investigation of  Self-supervised Task}
	To investigate the effectiveness of the hierarchical mutual information maximization (MIM), we break this procedure into two parts: local MIM between the user and user-centered sub-hypergraph, and global MIM between the user-centered sub-hypergraph and hypergraph. We then run MHCN with either of these two to observe the performance changes. We also compare hierarchical MIM with the node-graph MIM used in DGI to validate the rationality of our design. We implement DGI by referring to the original paper \cite{velickovic2019deep}. The results are illustrated in Fig. 7, and we use \emph{Disabled} to denote the vanilla MHCN. Unlike the bars in Fig. 6, each bar in Fig. 7 represents the case where only the corresponding module is used. As can be seen, hierarchical MIM shows the best performance while local MIM achieves the second best performance. By contrast, global MIM contributes less but it still shows better performance on Douban Yelp when compared with DGI. Actually, DGI almost rarely contributes on the latter two datasets and we can hardly find a proper parameter that can make it compatible with our task. On some metrics, training MHCN with DGI even lowers the performance. According to these results, we can draw a conclusion that the self-supervised task is effective and our intuition for hierarchical mutual information maximization is more reasonable compared with the node-graph MIM in DGI.

	\subsection{Parameter Sensitivity Analysis}
    In this section, we investigate the sensitivity of $\beta$ and $L$. \par
	As we adopt the $\emph{primary \& auxiliary}$ paradigm, to avoid the negative interference from the auxiliary task in gradient propagating, we can only choose small values for $\beta$. We search the proper value in a small interval and empirically set it from 0.001 to 0.5. We then start our attempts from 0.001, and proceed by gradually increasing the step size. Here we report the performance of $S^{2}$-MHCN with eight representative $\beta$ values $\{0.001, 0.005, 0.01, 0.02, 0.05, 0.1, 0.5\}$.  As can be seen in Fig. 8, with the increase of the value of $\beta$, the performance of $S^{2}$-MHCN on all the datasets rises. After reaching the peak when $\beta$  is 0.01 on all the datasets, it steadily declines. According to Fig. 8, we can draw a conclusion that even a very small $\beta$ can promote the recommendation task, while a larger $\beta$ would mislead it. The benefits brought by the self-supervised task could be easily neutralized and the recommendation task is sensitive to the magnitude of self-supervised task. So, choosing a small value is more likely to facilitate the primary task when there is little prior knowledge about the data distribution.\par
	
			\begin{figure}[t]
		\centering
		\includegraphics[width=0.5\textwidth]{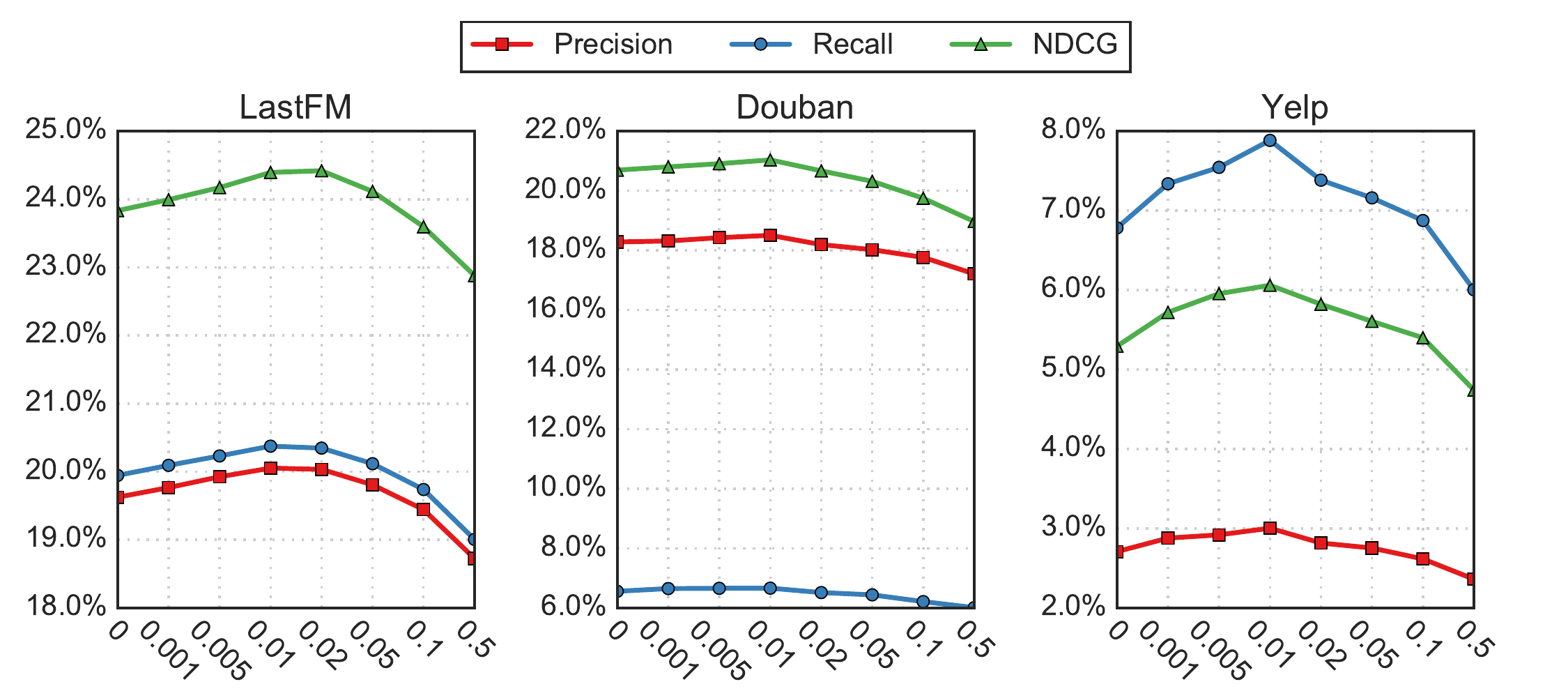}
		\caption{Influence of the magnitude of hierarchical MIM.}
		\label{figure.8}
	\end{figure}
	\begin{figure}[t]
		\centering
		\includegraphics[width=.5\textwidth]{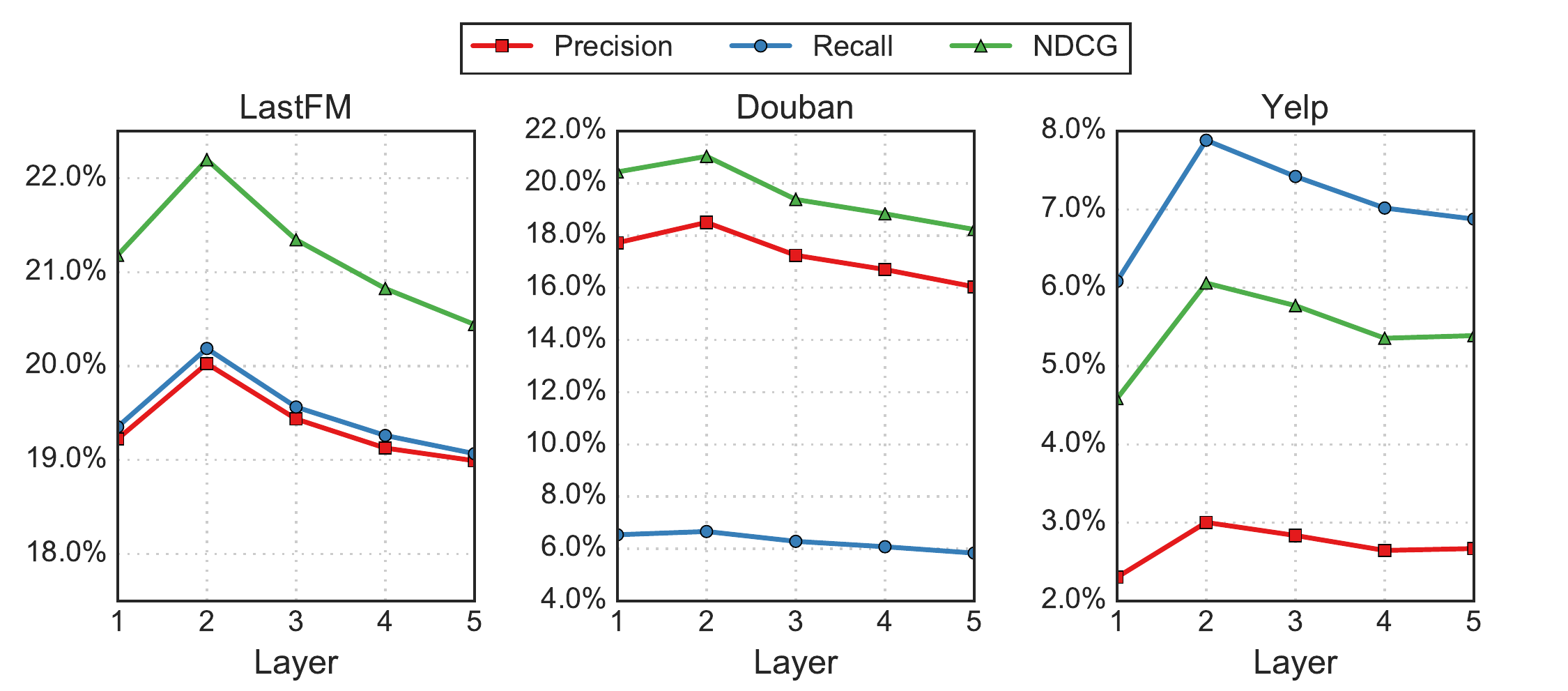}
		\caption{Influence of the depth of MHCN.}
		\label{figure.9}
		\vspace{-10px}
	\end{figure}

	Finally, we investigate the influence of $L$ to find the optimal depth for $S^{2}$-MHCN. We stack hypergraph convolutional layers from 1-layer to 5-layer setting. According to Fig. 9, the best performance of $S^{2}$-MHCN is achieved when the depth of $S^{2}$-MHCN is 2. With the continuing increase of the number of layer, the performance of $S^{2}$-MHCN declines on all the datasets. Obviously, a shallow structure fits $S^{2}$-MHCN more. A possible reason is that $S^{2}$-MHCN aggregates high-order information from distant neighbors. As a result, it is more prone to encounter the over-smoothing problem with the increase of depth. This problem is also found in DHCF \cite{ji2020dual}, which is based on hypergraph modeling as well. Considering the over-smoothed representations could be a pervasive problem in hypergraph convolutional network based models, we will work against it in the future. 

	\section{Conclusion}
	Recently, GNN-based recommendation models have achieved great success in social recommendation. However, these methods simply model the user relations in social recommender systems as pairwise interactions, and neglect that real-world user interactions can be high-order. Hypergraph provides a natural way to model high-order user relations, and its potential for social recommendation has not been fully exploited. In this paper,we fuse hypergraph modeling and graph neural networks and then propose a multi-channel hypergraph convolutional network (MHCN) which works on multiple motif-induced hypergraphs to improve social recommendation. To compensate for the aggregating loss in MHCN, we innovatively integrate self-supervised learning into the training of MHCN. The self-supervised task serves as the auxiliary task to improve the recommendation task by maximizing hierarchical mutual information between the user, user-centered sub-hypergraph, and hypergraph representations. The extensive experiments conducted on three public datasets verify the effectiveness of each component of MHCN, and also demonstrate its state-of-the-art performance.    
	
	\section*{Acknowledgment}
	This work was supported by ARC Discovery Project (Grant No. DP190101985 and DP170103954). Jundong Li is supported by National Science Foundation (NSF) under grant No. 2006844.
	
	\bibliographystyle{ACM-Reference-Format}
	\bibliography{refs}

\end{document}